\documentstyle{article}
\font\tenbb=msbm10
\font\sevenbb=msbm7
\font\fivebb=msbm5
\newfam\bbfam
\textfont\bbfam=\tenbb
\scriptfont\bbfam=\sevenbb
\scriptscriptfont\bbfam=\fivebb
\def\bb{\fam\bbfam}
\def\Cb{{\bb C}}
\newcommand{\be}{\begin{equation}}
\newcommand{\ee}{\end{equation}}
\newcommand{\bea}{\begin{eqnarray}}
\newcommand{\eea}{\end{eqnarray}}
\newcommand{\beas}{\begin{eqnarray*}}
\newcommand{\eeas}{\end{eqnarray*}}
\begin{document}
\title{Renormalization in Quantum Field Theory and the
Riemann--Hilbert problem}
\author{Alain Connes\thanks{connes@ihes.fr} \and Dirk
Kreimer\thanks{kreimer@ihes.fr}} \date{{\small {\em Institut des
Hautes Etudes Scientifiques}}\\ {\small Le Bois Marie, 35 route de
Chartres}\\ {\small F-91440 Bures-sur-Yvette, France}\\[3mm]
\small{\bf IHES/M/99/75}, hep-th/9909126} \maketitle

\begin{abstract}
\noindent We show that renormalization in quantum field theory is
a special instance of a general mathematical procedure of
multiplicative extraction of finite values based on the
Riemann--Hilbert problem. Given a loop $\gamma(z), \vert z
\vert=1$ of elements of a complex Lie group $G$ the general
procedure is given by evaluation of $ \gamma_{+}(z)$ at $z=0$
after performing the Birkhoff decomposition $
\gamma(z)=\gamma_{-}(z)^{-1} \gamma_{+}(z)$ where $
\gamma_{\pm}(z) \in G$ are loops holomorphic in the inner and
outer domains of the Riemann sphere (with $\gamma_{-}(\infty)=1$).
We show that, using dimensional regularization, the bare data in
quantum field theory delivers a loop (where $z$ is now the
deviation from 4 of the complex dimension) of elements of the
decorated Butcher group (obtained using the Milnor-Moore theorem
from the Kreimer Hopf algebra of renormalization) and that the
above general procedure delivers the renormalized physical theory
in the minimal substraction scheme.
\end{abstract}

\section{Introduction}
It has become increasingly clear in  \cite{K1} \cite{CK1}
\cite{K2} that the nitty-gritty of the perturbative expansion in
quantum field theory is hiding
 a beautiful underlying algebraic structure which does
not meet the eye at first sight.

As is well known most of the terms in the perturbative expansion are given
by divergent integrals
 which require renormalization. In \cite{K1} the renormalization technique
was shown to give
rise to a Hopf algebra whose antipode $S$ delivers the
same terms as those involved in the subtraction procedure
before the renormalization map $R$ is applied.
In \cite{CK1} the group $G$ associated to this Hopf algebra by the
Milnor--Moore theorem was computed by exhibiting a basis
and computing Lie brackets for its Lie algebra.\footnote{In the simplest
instance this group was later identified in \cite{Brouder}
as the Butcher group of numerical analysis.
For the general case we will use the terminology of decorated Butcher group.}
It was shown that the collection of all bare amplitudes
indexed by Feynman diagrams in dimensionally regularized perturbative
quantum field theory is just a point $\phi$
in the group $G_K$,
where $K=\Cb [z^{-1},[z]]$ is the field of Laurent series.

Though this made it clear that the Hopf algebra and its antipode are
providing the correct framework to understand renormalization,
some of the mystery
was still around because of the somewhat {\em ad hoc} manner,
in which the antipode $S$ had to be twisted by the renormalization map
$R$ in order to fully account for the physical computations.

The twisted antipode $S_R$ is defined recursively \cite{K1,K2} by
\begin{equation}
S_R(X)=-R[\phi(X)+S_R(X^\prime)\phi(X^{\prime\prime})],
\end{equation}
where we abbreviate the coproduct of $X$ as $\Delta(X)=X\otimes 1+1\otimes
X+X^\prime\otimes X^{\prime\prime}$,
omitting the summation sign for abbreviation always
and assuming $\bar{e}(X)=0$, where $\bar{e}$ is the counit.

The introduction of the twisted antipode is imposed by the actual
operations performed in renormalization. Thanks to
multiplicativity constraints (m.c.s), \[ R[XY]-R[R[X]Y]-R[X
R[Y]]+R[X]R[Y]=0,\] $S_R$ was shown in \cite{K2} to cover all
algebraic aspects of multiplicative renormalization,
$S_R(XY)=S_R(Y)S_R(X)$, including scale transformations and
changes of renormalization schemes by operations in the group.

But except from an obvious analogy between the recursive definition of $S_R$
 and the usual recursive construction of the antipode $S$ the conceptual understanding of
the twisted antipode was obviously missing.

We shall unveil the conceptual nature of $S_R$ thanks to the
Riemann--Hilbert problem \cite{RH} for the group $G$.

 This problem which
together with the inverse scattering method has been a very
successful tool to solve soliton equations, can be formulated as
follows: For a given analytic curve $C \subset \Cb P^1$ and a
 map $\gamma$ from $C$ to a complex Lie group $G$, find
the decomposition
\begin{equation}
\gamma(z)=\gamma_{-}(z)^{-1} \gamma_{+}(z),
\end{equation}
 of $\gamma$ as a pointwise product where $ \gamma_{+}(z)$ (resp.
$\gamma_{-}(z)$) is the boundary value of an holomorphic map from
the inner (resp. outer)
 domain of $C$ to the group $G$ and $ \gamma_{-}$ is normalized by
$\gamma_{-}(\infty)=1$. This decomposition is called the
Birk\-hoff decomposition.

What we shall show is that the renormalized theory is obtained
from the bare theory through the Riemann--Hilbert problem for the
group $G$ as follows.

As we have seen above, given a quantum field theory, the collection of all bare amplitudes
indexed by Feynman diagrams in dimensionally regularized perturbative
quantum field theory  is just a point $\phi$
in the group $G_K$,
where $K=\Cb [z^{-1},[z]]$ is the field of Laurent series.

In fact, looking more closely, this bare data is encapsulated as a
loop $\gamma(z)$ of elements in the group $G$, where $z$ is the
complex deviation from dimension four
 and lies in a small circle $C$ around the origin.

The renormalized theory is just the evaluation at $D=4$ of the
holomorphic piece  $\gamma_+$ in the Birk\-hoff decomposition of the
loop $\gamma$ as a product of two holomorphic maps $\gamma_{\pm}$
from the respective connected components $C_\pm$ of the complement
of the circle $C$ in the Riemann sphere $\Cb P^1$.

Thus, the nuance between naive subtraction of pole parts
and renormalization by local counterterms is the same
as passing from the additive Hilbert transform to
the multiplicative Riemann--Hilbert
problem.

This allows us to complete our understanding of the conceptual nature of renormalization
and to assert that, contrary to its reputation, the subtraction procedure
as applied by practitioners of QFT successfully over many decades
is now backed by its conceptual significance.

\section{The twisted antipode and the Birk\-hoff decomposition}
Let $H$ be a positively graded\footnote{We assume that the subspace $H_n$ of elements of degree $n$
  is finite dimensional for each $n$ and that $H_0$ is reduced
to scalars.}
 Hopf algebra over $\Cb$ which is commutative as an algebra.
Let $G$ be the group of characters of the algebra $H$, endowed with the group
operation $\phi_1\star\phi_2$
 given by the formula
\be
\phi_1\star\phi_2(X)=\phi_1 \otimes \phi_2( \Delta(X))
\;\, \, \forall
X\in H.
 \ee
where $\Delta$ is the coproduct.

 We shall show in this section that the Riemann--Hilbert problem
for $G$ yields exactly the defining equation for the twisted antipode $S_R$.

We let ${\cal R}$ be the ring of Laurent polynomials \footnote{The
discussion also applies to the ring of meromorphic functions in
$C_+$ whose only singularity in $C_+$ is a pole of finite order at
$0$.} and $R:\; {\cal R}\to {\cal R}$ be the linear projection on
the subalgebra ${\cal R}_-$ generated by $z^{-1}$ parallel to the
subalgebra ${\cal R}_+$ generated by $1,z$.

Let us first recall for clarity the standard dictionary between
the language of homomorphisms from $H\to {\cal R}$ (resp.~to
${\cal R}_{\pm}$) and the language of loops with values in $G$ and
domain the inner and outer components $C_{\pm}$ of the complement
of the unit circle $C$. We let ${\tilde H}$ be the augmentation
ideal of $H$, ie.~the kernel of the counit. The dictionary is then
the following: {\small
\begin{equation}
\begin{array}{rcl}
\underline{\mbox{{ Homomorphisms}  from $H\to {\cal R}$}} & \mid &
\underline{\mbox{{ Loops}  from $C$ to $G$}}\\
 & \mid & \\
 \phi({\tilde
H})\subset\;{\cal R}_- & \mid & \mbox{$\gamma$ extends to a
holomorphic map from $C_-\to G$}\\
 & \mid &
\mbox{with $\gamma(\infty)=1$.}\\ & \mid & \\ \phi({\tilde
H})\subset\;{\cal R}_+ & \mid & \mbox{$\gamma$ extends to a
holomorphic map from $C_+\to G$.}\\ & \mid & \\ \phi= \phi_1\star\phi_2
& \mid & \gamma(z)=\gamma_1(z)\gamma_2(z),\; \forall\; z\in {\bf
C}.\\ & \mid & \\
 \phi\circ S & \mid & z\to\gamma(z)^{-1}.
\end{array}
\end{equation}}

For elements $X \in{\tilde H}$ we shall use as above the short-hand notation
\be
\Delta(X)= X \otimes 1 + 1 \otimes X + X^\prime \otimes X^{\prime\prime}
\ee
where we omit the summation and the components $ X^\prime$ and $X^{\prime\prime}$
 are of degree strictly less than that of $X$ for $X \in H_n, n>0$.
\vspace{5mm}

 {\bf Theorem 1}: {\em
 Let $\phi$ be an homomorphism from $H\to {\cal R}$. The
$\phi_-$-component of the Birkhoff decomposition
$\phi=\phi^{-1}_-\star \phi_+$ of the corresponding loop is
characterized as the unique solution of the following equation}
\begin{equation}
\phi_-(X)=-R[\phi(X)+\phi_-(X^\prime)\phi(X^{\prime\prime})],\;\forall
X\in {\tilde H}.\label{sr}
\end{equation}

The main point of this theorem, of course, is that this equation is
identical to the defining equation
for the twisted antipode $S_R$.

Proof: By uniqueness of the solution of the Riemann--Hilbert
problem it is enough to exhibit one solution. Let us define
$\phi_-$ by $\phi_-   =\bar{e}$ (where $\bar{e}$ is the counit of
$H$) on elements of $H$ of degree zero and then by induction,
using the above equation. The first thing to check which is not
obvious is that this definition actually yields a homomorphism.
This follows from \cite{K2} using the multiplicativity property of
the map $R$
\be
R[xy]-R[R[x]y]-R[xR[y]]+R[x]R[y]=0.\label{mc} \ee For the sake of
completeness we briefly recall the main ingredients of the proof.
It suffices to prove the assertion for homogeneous elements $X,Y$
of non-zero degree. This is done by induction on the sum of the
respective degrees. Writing the recursive definition for both
$\phi_-(X)$ and $\phi_-(Y)$, and taking the product yields, using
(\ref{mc}) a sum of terms which one has to equate with the terms
coming from the recursive definition applied to $XY$. One has
\beas \Delta(XY) & = & XY\otimes 1+1\otimes XY +X\otimes
Y+Y\otimes X+ X^\prime Y\otimes X^{\prime\prime} + X^\prime\otimes
X^{\prime\prime}Y\\
 & &
+\;X Y^\prime \otimes Y^{\prime\prime}+ Y^\prime \otimes
XY^{\prime\prime}+ X^\prime Y^\prime \otimes X^{\prime\prime}
Y^{\prime\prime}. \eeas Using the induction hypothesis, i.e.~the
multiplicativity of $\phi_-$ together with the defining equation
of $\phi_-$ one easily reorganizes the above nine terms, using (6),
to get the
desired result.

Now that we know that $\phi_-$ is a homomorphism we just need to
check that $\phi_-({\tilde H})\subset{\cal R}_-$ which is obvious by
construction since ${\cal R}_-$ is the range of the projection $R$,
 and that the product $\phi_-\star \phi$ lands in
${\cal R}_+$. But ${\cal R}_+$ is the kernel of $R$ and it
suffices to check that the composition $R[\phi_-\star \phi]$
vanishes. But this follows directly from the defining equation
using $R[\phi_-]=\phi_-$. $\Box$
\section{Multiplicative Renormalization}
Let us go back to the twisted antipode $S_R$. First of all, $S_R$
should be viewed as a homomorphism from the Hopf algebra $H$ to
the ring of meromorphic functions in $C_+$ whose only singularity
 in $C_+$ is a pole of finite order at $0$.
Its recursive definition starts with the given homomorphism $\phi$
from $H$ to $K$ and modifies it by induction using the above
formula \begin{equation}
S_R(X)=-R[\phi(X)+S_R(X^\prime)\phi(X^{\prime\prime})],
\end{equation}
while $S_R(e)=1$.
\vspace{5mm}
\\ {\bf Theorem 2}: {\em  The transition from the
bare Green function $\phi(X)$ to the MS-renormalized Green
function $\psi(X)=[S_R\star\phi(X)]$ is the multiplicative
projection of the bare Green function $\phi$ to the holomorphic
part $\phi_+$ followed by evaluation at the origin.}\footnote{This
result is applicable as long as the theory under consideration
allows for local counterterms, which includes renormalizable
theories but also effective theories.}
\vspace{5mm}

The proof of this theorem relies on the
results of \cite{K1}\cite{CK1}\cite{K2},
on theorem 1 and on the detailed
discussion of dimensional regularization in the next section.

The relation between different renormalization schemes
explained in \cite{K2} also explains the finiteness of Green
functions when we vary dimensionful
parameters. It is  common practice to maintain
dimensionless coupling constants in dimensional regularization by
introducing a dimensionful parameter $[\mu^2]^{z}$ into the
integral measure, as in the example exhibited in the next section.
The Birkhoff decomposition is done at $q^2=\mu^2$,
say.
At other $q^2$, one imposes on
 Feynman graphs $\Gamma$ with $deg(\Gamma)=n$ loops  a
homomorphism $\rho(\Gamma)=[\mu^2/q^2]^{z\;deg(\Gamma)}$
which is in accordance with the m.c.s.

It is important to stress that in the above multiplicative
renormalization procedure
 the precise location of the curve C is not crucial since one can move
it freely by a homotopy in the complement of the singularities of
the initial loop $\gamma$.

We shall end the paper with a few  comments on the known
crucial features
of dimensional regularization which enabled us to obtain this
result.
\section{Dimensional regularization}
Dimensional regularization (Dim.Reg.) can be characterized by the three
following properties: it naturally involves $\Cb P^1$, it projects
to  logarithmic divergences and it raises internal propagators to
complex powers.

It promotes the dimension four of spacetime consistently
to an analytic continuation
to $D=4-2z$, $z\in \Cb$, so that the complex parameter $z$ serves
as a regularization parameter. It is not necessary to exhibit the
technical
details of this analytic continuation, we rather comment
on its most useful properties mentioned above.

All the above three properties distinguish it from regularization
prescriptions which use a dimensionful regularization parameter,
like in a cut-off or  Pauli-Villars scheme, where
\begin{itemize}
\item a dimensionful parameter $\lambda$ parametrizes the divergences
by a finite series in $\lambda$ and $\log\lambda$, and the
convergent part as a series in $1/\lambda$, spoiling any attempt
towards using $\lambda$ in a holomorphic decomposition of the
finite and divergent part;
\item it involves no projection onto logarithmic divergences at the
level of the regularization;
\item integration of subgraphs evaluates
to logarithms of internal propagators instead of complex powers.
\end{itemize}

The practical advantages of Dim.Reg.~in this respect are so
severe\footnote{See \cite{David} for a typical example of a
calculation which would have been barely possible without
Dim.Reg.~and the multiplicativity properties of counterterms.}
that even a prescription like on-shell BPHZ, fully avoiding
regularization at all, cannot compete with Dim.Reg.~in practical
calculations. We now see that this is underwritten from a
conceptual viewpoint: it is actually the presence of complex
domains provided by $z=(4-D)/2\in\;\Cb$ in Dim.Reg.~which allows
to promote renormalization to a concept.

Let us for the sake of the reader exhibit these features in some detail
using simple
examples. Let us consider the basic integral
in dimensionally regularized Euclidean space
\[
\int \frac{{\rm d}^Dk}{[\mu^2]^{-z}} \frac{1}{k^2+m^2}=-m^2\int
\frac{{\rm d}^Dk}{[\mu^2]^{-z}} \frac{1}{k^2(k^2+m^2)}
+\int \frac{{\rm d}^Dk}{[\mu^2]^{-z}} \frac{1}{k^2}.\label{ex1}
\]
In Dim.Reg., the last expression on the rhs is zero thanks to analytic
continuation, which gives \cite{Collins}, as one of Dim.Reg.s
defining properties,
\[\int \frac{{\rm d}^Dk}{[\mu^2]^{-z}} [k^2]^a=0,\]
for all $a\in \Cb$.
The first expression on
the rhs of (\ref{ex1})
evaluates to $-m^2[m^2/\mu^2]^{-z}$ $\pi^{2-z}\Gamma(-1+z)$,
with the $\Gamma(-1+z)$-function delivering the expected pole term near
$z=0$.

Evaluating the same quadratic divergent
integrand by a standard integral in
 four dimension, multiplying the integrand with a cut-off function
$\Theta[\lambda-\sqrt(k^2)]$, we find the result
$
-m^2\pi^2[\log\lambda^2/m^2+\log(1+m^2/\lambda^2)]+\pi^2\lambda^2,
$
where $\lambda\to\infty$ is now the limit of interest.

What can we learn from this basic example? First of all, we use it
to exhibit what we mean when we say that Dim.Reg.~naturally
involves $\Cb P^1$. Clearly, the bare result in Dim.Reg.~is a
series which has a pole in $z$ of finite order (first order in the
example) and is an infinite Taylor series in $z$. We hence can
formulate the quest for the Birkhoff decomposition in the
framework used above.

Further, the result $\int \frac{{\rm d}^Dk}{[\mu^2]^{-z}}
[k^2]^a=0$ immediately ensures that one only confronts logarithmic
divergent integrals, with suitable dimensionful coefficients
($-m^2$ in our example) maintaining the correctness of
powercounting. This has far reaching consequences: it allows to
add zero in a suitable manner to each Feynman integrand to
dispense with all appearances of overlapping divergences, as
exemplified by the following instructive example:{\small
\[
\int  \frac{d^Dk}{[\mu^2]^{-z}}\frac{d^Dl}{[\mu^2]^{-z}}
\frac{1}{(k^2+m^2)(k+l)^2l^2}= \int
\frac{d^Dk}{[\mu^2]^{-z}}\frac{d^Dl}{[\mu^2]^{-z}} \left[
\frac{-m^2}{(k^2+m^2)k^2(k+l)^2l^2} \right], \]} where the lhs and
rhs are equal using an addition of {\small
\[
0=-\int \int \frac{d^Dk}{[\mu^2]^{-z}}\frac{d^Dl}{[\mu^2]^{-z}}
\frac{1}{k^2(k+l)^2l^2}.\]} In a regularization using a
dimensionful parameter, such a simplification would not appear
until one uses the Hopf algebra structure to realize that terms
which do not depend on dimensionful parameters drop out at the end
of the calculation in ratios of the form $S_R\star \phi$. Hence,
in Dim.Reg.~we already obtain at the level of bare diagrams a
decomposition of amplitudes into functions representing rooted
trees, hence the desired representation in terms of elements of
the decorated Butcher group $G_K$.
\section*{Acknowledgements}
Both of us are grateful to the Clay Mathematics Institute
and to Institut des Hautes Etudes Scientifiques
for making our collaboration possible.
D.K.~thanks the DFG for a Heisenberg Fellowship.


\begin{thebibliography}{99}
\bibitem{K1}
D.~Kreimer, Adv.Theor.Math.Phys.{\bf 2} (1998) 303,
q-alg/9707029.
\bibitem{CK1}
A.~Connes, D.~Kreimer, Commun.Math.Phys.{\bf 199} (1998) 203,
hep-th/9808042.
\bibitem{K2}
D.~Kreimer, Adv.Theor.Math.Phys.{\bf 3.3} (1999)
hep-th/9901099.
\bibitem{Brouder} C.~Brouder, {\em Runge-Kutta methods and renormalization},
hep-th/9904014.
\bibitem{RH}
Math\'ematique et Physique, (S\'eminaire de l'ENS 79-82). Progress
in Math. 37, Birkh\"auser, Boston (1983).
\bibitem{Collins} J.C.~Collins, {\em Renormalization}, Cambridge Univ.Press
1984.
\bibitem{David}
D.J.~Broadhurst, {\em Four loop Dyson--Schwinger--Johnson
anatomy}, to appear in Phys.Lett.{\bf B}, hep-ph/9909336.
\end{thebibliography}
\end{document}